\documentclass[graybox]{svmult}
\usepackage{mathptmx}       
\usepackage{helvet}         
\usepackage{courier}        
\usepackage{type1cm}        
\usepackage{graphicx}
\usepackage{array}
\usepackage{newtxtext}       %
\usepackage[varvw]{newtxmath}       

\usepackage{amsmath} 
\usepackage[style=apa,backend=biber]{biblatex}
\newcommand{\bfs}{\boldsymbol} 

\usepackage{comment}
\usepackage{graphicx}
\usepackage{booktabs}
\usepackage{multirow}
\usepackage{ragged2e}
\usepackage{lineno}
\RequirePackage[colorlinks,citecolor=blue,urlcolor=blue]{hyperref}

\begin{document}

\title{Assessment of Misspecification in CDMs using a Generalized Information Matrix Test}
\author{Reyhaneh Hosseinpourkhoshkbari \href{https://orcid.org/0009-0000-9638-6814}{\includegraphics[scale=0.5]{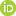}} and Richard M. Golden \href{https://orcid.org/0000-0001-7505-6832}{\includegraphics[scale=0.5]{ORCIDiD.png}}}
 \authorrunning{Hosseinpour and Golden}

\institute{Reyhaneh Hosseinpourkhoshkbari
\at Cognitive Informatics and Statistics Lab, 
School of Behavioral and Brain Sciences (GR4.302), \\ University of Texas at Dallas, Richardson, TX.
\email{reyhaneh.hosseinpour@utdallas.edu}
\and Richard M. Golden 
\at Cognitive Informatics and Statistics Lab, School of Behavioral and Brain Sciences (GR4.1), \\
University of Texas at Dallas, Richardson, TX.
 \email{golden@utdallas.edu}\\ \\
This project was partially funded by The University of Texas at Dallas Office of Research and Innovation by an award to Richard Golden
through the SPARK program.}
%
%
\maketitle
\abstract*{If the probability model is correctly specified, then we can estimate the covariance matrix of the asymptotic maximum likelihood estimate distribution using either the first or second derivatives of the likelihood function. Therefore, if the determinants of these two different covariance matrix estimation formulas differ this indicates model misspecification. This misspecification detection strategy is the basis of the Determinant Information Matrix Test ($GIMT_{Det}$). To investigate the performance of the $GIMT_{Det}$, a Deterministic Input Noisy And gate (DINA) Cognitive Diagnostic Model (CDM) was fit to the Fraction-Subtraction dataset. Next, various misspecified versions of the original DINA CDM were fit to bootstrap data sets generated by sampling from the original fitted DINA CDM. The $GIMT_{Det}$ showed good discrimination performance for larger levels of misspecification.
In addition, the $GIMT_{Det}$ did not detect model misspecification when model misspecification was not present
and additionally did not detect model misspecification when the
level of misspecification was very low. However, the $GIMT_{Det}$ discrimation performance was highly variable across different misspecification strategies when the misspecification level was moderately sized. The proposed new misspecification detection methodology is promising but additional empirical studies are required to further characterize its 
strengths and limitations.}

\abstract{If the probability model is correctly specified, then we can estimate the covariance matrix of the asymptotic maximum likelihood estimate distribution using either the first or second derivatives of the likelihood function. Therefore, if the determinants of these two different covariance matrix estimation formulas differ this indicates model misspecification. This misspecification detection strategy is the basis of the Determinant Information Matrix Test ($GIMT_{Det}$). To investigate the performance of the $GIMT_{Det}$, a Deterministic Input Noisy And gate (DINA) Cognitive Diagnostic Model (CDM) was fit to the Fraction-Subtraction dataset. Next, various misspecified versions of the original DINA CDM were fit to bootstrap data sets generated by sampling from the original fitted DINA CDM. The $GIMT_{Det}$ showed good discrimination performance for larger levels of misspecification.
In addition, the $GIMT_{Det}$ did not detect model misspecification when model misspecification was not present
and additionally did not detect model misspecification when the
level of misspecification was very low. However, the $GIMT_{Det}$ discrimation performance was highly variable across different misspecification strategies when the misspecification level was moderately sized. The proposed new misspecification detection methodology is promising but additional empirical studies are required to further characterize its 
strengths and limitations.}

\section{Introduction}
{\label{154334}}
Cognitive diagnostic models (CDMs) are a family of restricted latent class psychometric models designed to assess examinee skill mastery by incorporating prior knowledge of the relationship between latent skills and student responses (\cite{Torre2009}). 
In addition, a CDM outputs the attribute distribution or the probability that an examinee will have a particular set of skills given the examinee's exam performance.
Characteristics of latent skills in CDMs are defined through a Q-matrix that specifies which specific skills are relevant for answering a particular exam question. Furthermore, Q-matrix misspecification can affect parameter estimates and respondent classification accuracy (\cite{RuppTemplin2008}). Therefore, careful validation of the Q-matrix is crucial to ensure that the model accurately represents the underlying relationships between the test items and the measured attributes. Methods have been developed to estimate and validate expert knowledge. However, despite best intentions, the possibility of CDM misspecification is always present.

The effects of model misspecification in CDMs have been investigated by directly comparing observed frequencies and predicted probabilities (\cite{Kunina2012}), by comparing a correctly specified model to a nested model using Wald test methodology (\cite{Torre2011}), and comparing selected first and second order observed and expected moments (\cite{Chen2018}).  These methodologies often face the challenge
of test statistics with poor statistical power due to the difficulty of 
reliably estimating parameters in fully saturated (highly flexible models).
For example, consider the problem of comparing the predicted probability of 
every pattern of responses to an exam with $d$ items using a probability
model with $q$ free parameters. The degrees of freedom (and variance) for the Pearson  Goodness-Of-Fit (GOF) chi-squared
test statistic increases as an exponential
function of the number of items $d$. The M2 statistic
(\cite{maydeu2014})
provides an improvement
in statistical power by only examining the first and second moments resulting
in a chi-squared test statistic whose degrees of freedom increase as a quadratic
function of the number of items $d$.

Using a different approach, which detects model misspecification by comparing
different covariance matrix estimators,
\textcite{Wh82} introduced the Information Matrix Test (IMT) which is a chi-squared test with $q(q-1)/2$ degrees of freedom. This test is based on comparing the inverse covariance matrix estimators: $\hat{{\bf A}}^{-1}$, derived from the Hessian matrix, and $\hat{{\bf B}}^{-1}$, derived from the outer product gradient (OPG). These estimators are calculated from the second and first derivatives of the log-likelihood function, respectively. The \textcite{Wh82} IMT test statistic for
misspecification detection has a chi-squared distribution whose degrees
of freedom increase as a quadratic function of the number of parameters $q$
in the model. \textcite{golden2013,Golden2016} proposed a generalization of 
the \textcite{Wh82} IMT framework called the Generalized Information Matrix Test (GIMT)
framework. However, no systematic studies have been conducted to specifically investigate the performance of the GIMT in the context of model misspecification detection in CDMs.

\textcite{presnell2004ios} developed and empirically evaluated
a statistical test for comparing the "in-sample" (training data)
log-likelihood model fit to the "out-of-sample" (test data)
log-likelihood model fit. Presnell and Boos (2004) referred to
their test as the "in and out of sample" (IOS) test and showed
the test statistic had a chi-squared distribution with 1 degree
of freedom regardless of model or data complexity.
The IOS test may be interpreted as a type of
GIMT as described by \textcite{golden2013, Golden2016} which we 
call the {\em Determinant GIMT} ($GIMT_{Det}$) $GIMT_{Det}=(1/q) \log\det(\hat{\bf A}_n ^{-1}\hat{\bf B}_n)$. 

More recently, \textcite{liu2019} examined the performance of different methods in providing consistent standard errors (SEs) for item parameter estimates for situations where a CDM was misspecified or correctly specified. \textcite{liu2019} showed a difference among OPG, Hessian, and Robust standard errors when the Q-matrix is misspecified. Although \textcite{liu2019} had not intended to develop a method for misspecification
detection, their empirical results nevertheless support a type of GIMT for misspecification detection as introduced by \textcite{golden2013, Golden2016}. 

In this paper,
we describe another GIMT methodology that focuses on a single statistic
for comparing covariance matrices rather than a comparison based upon 
their diagonal elements. First, we sketch the mathematical
derivation of the asymptotic distribution of the $GIMT_{Det}$
statistic for CDMs
using the methods described by 
\textcite{golden2013, Golden2016}. Second, we empirically investigate the asymptotic
behavior of the $GIMT_{Det}$ for CDMs in a series of simulation
studies.
The simulation studies simulate data sets from a known DINA CDM fit to the \textcite{Ta84} Fraction-Subtraction data set. Next, the resulting bootstrapped simulated data sets are fit to the DINA CDM which generated the bootstrap data as well as different misspecified versions of the original DINA CDM. The discrimination
performance of the $GIMT_{Det}$ is then reported to provide
an empirical evaluation of the ability of the $GIMT_{Det}$ to detect model misspecification in a DINA CDM. 

\section{Mathematical Theory}
\subsection{Model Misspecification}
In the statistical machine learning framework, the Data Generating Process (DGP) generates observable data directly from an underlying unobservable probability distribution, $p_{\text{DGP}}$.  A probability model,$M$, is a collection of probability mass functions. If $p_{DGP} \in M$, then $M$ is correctly specified; otherwise, $M$ is misspecified with respect to $p_{\text{DGP}}$. Less formally, a model capable of
representing the DGP is called a "correctly specified model." 
\subsection{{Cognitive Diagnostic Model Specification~}}

\subsubsection{Data Set} 
Consider a scenario where $N$ examinees are randomly selected to take a diagnostic test of $J$ items. The outcomes of this test are recorded in a binary matrix $\bf X$, which has $N$ rows and $J$ columns. Each element, ${ x}_{ij}$, represents the response of examinee $i$ $(1\leq i \leq N)$ to item $j$ $(1\leq i \leq N)$, with a value of $1$ indicating a correct response and $0$ indicating an incorrect response. The rows of the matrix  {\bf X},
${\bf x}_1, \ldots, {\bf x}_n$,
are assumed to be a realization of a sequence of independent and identically distributed random vectors with a common probability mass function $p_{DGP}(x)$.

\subsubsection{Evidence Model}
Let ${\bf \alpha} = (\alpha_{1}, \ldots, \alpha_{k})$
 represent the discrete attribute mastery profile for an examinee, where  $\alpha_{k}$ is one if and only if that examinee demonstrates mastery of latent skill $k$. Let ${ \bf q}_j=[q_{j1}, \ldots , q_{jk}]$ represent the $j$th row of the $\bf Q$-matrix, where each element, $q_{jk}$ equals one if the $k$th skill is required to answer question $j$ and is equal to zero otherwise.
Let logistic sigmoidal function ${\cal S}(\phi)$ be defined such that ${\cal S}(\phi) = 1/(1 + \exp(-\phi))$.  Let ${\bfs \beta} =[\bfs \beta_1, .., \bfs \beta_j]^T$ denote item parameters vector, where ${\bfs \beta_j}= [\beta_{j1}, \beta_{j2}]^T$ and $\beta_{j1}$ is the main and interaction effect parameters for question $j$, and $\beta_{j2}$ is the intercept parameter. The probability of a correct response to question $j$, given the mastery profile, item parameters, and the skills required for question $j$, specified by ${\bf q}_j$, is calculated using the following formula:
\begin{displaymath}
p_{ij}=p\left ({x}_{ij} = 1 | { \bfs \alpha}, {  \bfs\beta}_j, { \bfs q}_j \right)
= {\cal S}({ \bfs \beta}_{j1} {\psi}({ \bfs \alpha}, { \bf q}_j)-{ \bfs \beta}_{j2}).
\end{displaymath}
where larger values of $\psi(\alpha, {\bf q}_j)$ indicate
an increased likelihood that item $j$
is correctly answered given
that latent skill pattern $\alpha$. We have implemented the DINA CDM by defining 
${\psi}({ \bfs \alpha}, { \bfs q}_j) = 
 (\prod_{k=1}^K  {\bfs \alpha}_{k}^{q_{j,k}}) \in \{0,1\}$
to calculate the expected response to item $j$. \\
Let:
\begin{displaymath}
p\left (x_{ij}| { \bfs\alpha}, { \bfs\beta}, { \bfs q}_j \right)= x_{ij} p_{ij}+(1-x_{ij} )(1-p_{ij}).
\end{displaymath}
The probability of all observed responses for $i$th examinee, $x_i$, given a specific pattern of latent skills, can be expressed as follows:

\begin{displaymath}
p({ x}_i | { \bfs \alpha}, { \bfs \beta_j}) =
\prod_{j=1}^J p(x_{ij} | { \bfs\alpha}, { \bfs \beta_j}, { \bfs q}_j).
\end{displaymath}

\subsubsection{Proficiency Model}
After specifying the conditional distribution of student responses given latent skill profiles,  the subsequent emphasis will be on investigating the joint distribution of attributes. The saturated joint attribute distribution model for all possible values that 
the $k$-dimensional binary vector ${\bfs \alpha}$ can take on requires $2^k-1$ parameters, so when the number of attributes is moderately large, a more constrained joint attribute model might be desired to represent the joint distribution of ${\bfs \alpha}$. In this paper, a Bernoulli latent skill attribute probability model (e.g., \cite{Maris1999}) is assumed where the latent skills are
independent so that the probability that the $k$th latent skill, $\alpha_k$, is present in the attribute pattern is 
given by the formula:
\begin{displaymath}
p(\alpha_k ) = \alpha_k {\cal S}(-\eta_k)
+ (1 - \alpha_k)(1 - {\cal S}(-\eta_k))
\end{displaymath}
where $\eta_k$ can be a free parameter; however, in this paper, $\eta_k$ is a constant chosen such that the guess probability ${\cal S}(-\bfs \eta$) is 0.354.
The probability, $p({\bfs \alpha}|\bfs \eta)$, of a skill attribute profile, 
${{\bfs \alpha}} = [\alpha_{1}, \ldots, \alpha_{K}]$,
for an examinee given $\eta$ is given by the formula:
\begin{displaymath}
p({{\bfs \alpha}}) = \prod_{k=1}^K p(\alpha_{k} ).
\end{displaymath}

\subsection{{Model Parameter Estimation~}}

The parameter prior for the two-dimensional parameter vector $\bfs \beta_j$, associated with the $j$th question, is represented by a bivariate Gaussian density, denoted as $p(\bfs \beta_j)$. This density has a two-dimensional mean vector $\bfs \mu_j$ and a two-dimensional covariance matrix $\bf C_{\bfs \beta}$ for $j = 1, \ldots, J$. It is assumed that the constants $\mu_1, \ldots, \mu_J$ are known, and $\bfs C_{\bfs \beta}$ is a positive number. Let $\bfs \mu_\beta = [\mu_1, \ldots, \mu_J]$. The joint distribution of $\bfs \beta = [\bfs \beta_1, \ldots, \bfs \beta_J]$ is specified as the \textit{parameter prior}: 
\begin{equation}
    p(\bfs \beta) = \prod_{j=1}^{J} p({\bfs \beta}_j).
    \label{prior}
\end{equation}

{\label{732142}}
The {\em likelihood function} of the response vector of examinee $i$ who is assumed to have attribute pattern ${\bfs \alpha}$ is given by:

\begin{equation}
\label{likelihood}
p({\bf x}_i , {\bfs \alpha} | {\bfs \beta}) = p({\bfs \alpha})p({\bf x}_i | {\bfs \alpha}, {\bfs \beta}) =p({\bfs \alpha}) \prod_{j=1}^J 
p(x_{ij} | {\bfs \alpha}, {\bfs \beta}_j, {\bf q}_j).
\end{equation}

yielding the MAP empirical risk function:
\begin{equation}
\label{MAPriskloss}
\hat{\ell}_n({\bfs \beta}) 
= -(1/n) \log p({\bfs \beta}) + (1/n)\sum_{i=1}^{n} c({\bf x}_i, {\bfs \beta}).
\end{equation}
where
\begin{displaymath}
\;\; c({\bf x}_i; {\bfs \beta}) = -\log p({\bf x}_i | {\bfs \beta}) = - \log 
\sum_{\alpha}
p({\bf x}_i | {\bfs \alpha} , {\bfs \beta}) p({\bfs \alpha}).
\end{displaymath}

In this study, the MAP empirical risk function (3) involves summing over all possible latent skill attribute patterns.
 Assume $\hat{\ell}_n$ is a twice-continuously differentiable objective function.
Once a critical
point is reached, then the Hessian of $\hat{\ell}_n$ can be evaluated
at that point to check if the critical point is a strict local
minimizer (e.g., \cite{Golden2020}).
Here we assume that
a parameter estimate $\hat{ \bfs \beta}_n$ has been obtained which
is a strict local minimizer of $\hat{\ell}_n({\bfs \beta})$ 
in some (possibly very small) closed, bounded, and convex
region of the parameter space $\Omega$ for all sufficiently large
$n$. Furthermore, we assume that
the expected value of $\hat{\ell}_n({\bfs  \beta})$, 
$\ell({\bfs  \beta})$, has a strict global minimizer,
${\bfs \beta}^*$, in the interior of $\Omega$.
This setup of the problem thus allows for situations where
$\ell$ has multiple minimizers, maximizers, and saddlepoints over the
entire unrestricted parameter space.
Given these assumptions, it can be shown (e.g., \cite{Wh82,Golden2020}) that $\hat{ \bfs \beta}_n$ in this case is a
consistent estimator of ${\bfs \beta}^*$. In addition, for large sample sizes, the effects of the parameter prior becomes negligible and the MAP estimate has the same asymptotic distribution as the maximum likelihood estimate.
\subsection{Information Matrix Test Methods for Detection of Model Misspecification}
 
\subsubsection{Determinant  Generalized  Information 
Matrix Test Statistical Theory}
In this section, we present explicit details regarding
the derivation of the asymptotic distribution of
the $GIMT_{Det}$. 
Let ${\bf g}({\bf \tilde{x}};\bfs \beta) \equiv - \nabla \log p ({\bf \tilde{x}}_i; {\bfs \beta})$. 
Let ${\bf A}^*$ denote the Hessian of $\ell$, evaluated 
at ${ \beta}^*$,
 ${\bf A}^*={\bf A}({ \bfs \beta^*})=  -\nabla^2 E \{ \log p (({\bf \tilde{x}}_i; {\bfs \beta}) \}.$
Let ${\bf B}^*$ denote 
${\bf B^*} ={\bf B}({ \bfs \beta^*}) = E \{ {\bf g}({\bf \tilde{x}};\bfs \beta) {\bf g}({\bf \tilde{x}};\bfs \beta)^T\}$
evaluated at the point ${\bfs \beta}^*$.
It is well known (e.g., \cite{Wh82,golden2013,Golden2016}) that if ${\bf A}^*$ and
${\bf B}^*$ are positive definite, then the asymptotic distribution of $\hat{\bfs \beta}_n$
is a multivariate Gaussian with mean ${\bfs \beta}^*$ and covariance matrix 
$(1/n) {\bf C}^* \equiv (1/n) ({\bf A}^*)^{-1} {\bf B}^* ({\bf A}^*)^{-1}$ as $n \to \infty$.
In the special case where the model $M$ is correctly specified in the sense that
the observed data is i.i.d. with a common probability mass function such that:
$p( {\bf x} | {\bfs \beta}^*) = p_{DGP}({\bf x})$, then
the covariance matrix ${\bf C}^*$ can be computed using either
$({\bf A}^*)^{-1}$ or $({\bf B}^*)^{-1}$. It then follows that if $M$ is correctly specified with respect to $p_{DGP}$, then ${\bf A}^*$ = ${\bf B}^*$.
Consequently, if ${\bf A}^* \neq {\bf B}^*$, then $M$ is misspecified with respect to $p_{DGP}.$

Let {$ s: {\cal R}^{q \times q} \times {\cal R}^{q \times q} \rightarrow {\cal R}
$ } be a continuously differentiable function and 
assume $\nabla s$ evaluated 
at ${\bf A}^*$ and ${\bf B}^*$ has full row rank. The function $s$ is called a GIMT Hypothesis Function that has the property that if ${\bf A}$ = ${\bf B}$, then ${ s(\bf A,\bf B)= 0}$ for every symmetric positive definite matrix {\bf A} and for every symmetric positive definite matrix {\bf B}. Let ${s^*\equiv s({\bf A}^*,{\bf B}^*)}$ following \textcite{golden2013, Golden2016}.

Let $\hat{\bf A}_n$ denote the Hessian of $\hat{\ell}_n$ evaluated
at $\hat{\bfs \beta}_n$. Let
\begin{displaymath}
\hat{\bf B}_n = (1/n) \sum_{i=1}^n {\bf g}(\tilde{\bf x}_i, \hat{\bfs \beta}_n){\bf g}(\tilde{\bf x}_i, \hat{\bfs \beta}_n)^T.
\end{displaymath}
A GIMT is defined as a test statistic $\hat{s}_n \equiv s({ \hat{\bf A}_n},{\hat{\bf B}_n})$  that 
evaluates  the  null  hypothesis: 
\begin{displaymath}
    {H_0 : s^* = { 0}}.
\end{displaymath}

The $GIMT_{Det}$ (\cite{golden2013}; also see \cite{Golden2020, Golden2016}) is specified by the GIMT hypothesis function $s$ is defined such that: 
$ s({\bf {A^*}},{\bf {B^*}}) = \frac{1}{\textit{q}} \log \det(\bf (A^*)^{-1}\bf B^*).$
Let the Wald test statistic: 
$${\bf \mathcal{W}}_{\text{n}} = n( \hat{s}_n)^T \hat{\bf C}_{n,s}^{-1}( \hat{s}_n).$$
where $ \hat{s}_n = \frac{1}{q} \log \det((\bf \hat{A})^{-1}\bf \hat{B})$, 
 If the null hypothesis ${H_0: s^* = 0}$ holds, then ${\bf \hat{\mathcal{W}}}_n$ converges in distribution to a chi-squared random variable with one degree of freedom as $n \rightarrow \infty$ (see Theorem 7 for \cite{Golden2016}). 
 If the null hypothesis fails, ${\bf\hat{\mathcal{W}}}_n \rightarrow \infty$ as $n \rightarrow \infty$ with a probability of one (see Theorem 7 for \cite{Golden2016}).
Here the estimator $ \hat{\bf C}_{n,s}^{-1}$ is an estimate of the asymptotic covariance matrix of $n^{1/2} ({ \hat{s}_n}-s^*)$, $({\bf C}_{s}^{*})^{-1}$.
\cite{golden2013, Golden2016} shows how this asymptotic covariance matrix can be estimated using the first, second, and third derivatives
of the log-likelihood function. Notice that since the degrees of freedom of the Wald statistic for the $GIMT_{Det}$ is equal to 1, the test statistic has a distribution that is the square of a normalized Gaussian random variable.

\section{Simulation Study} 
\subsection{Dataset}
The simulation studies used CDMs with parameters estimated from the
\textcite{Ta84} fraction-subtraction data set. The dataset consists of a set of math problems involving fractions subtraction, designed to assess students' ability to solve problems involving fractions. In this case, we utilized the 'Fraction.1' dataset from the 'CDM' R package (\cite{cdmR}) containing the dichotomous responses of 536 middle school students over 15 fraction subtraction test items. The $\bf Q$-matrix specifying which of five distinct skills were required to answer a particular test item was based upon the $\bf Q$-matrix used by \textcite{Ta84} and provided in \textcite{cdmR}.

\subsection{Methods}
First, the described DINA CDM was fit to the data set with a sample size of $n=536$ using the previously described MAP estimation method (see Equation \ref{MAPriskloss}). A multivariate Gaussian prior with mean vector $\bfs \mu_{\bfs \beta}$ and covariance matrix $\bf C_{\bfs \beta}$ was used for all items (see Equation \ref{prior}). To ensure guess (g) and slip (s) probabilities both equaled 0.354 for a specific item, the Gaussian Parameter Prior mean vector for that item was set as 
$ \bfs \mu_{\bfs \beta}= \begin{bmatrix}
    1.2 &  0.6
\end{bmatrix}^T$. An uninformative Gaussian Parameter Prior covariance matrix $\bf C_{\bfs \beta}$ was chosen to be $\sigma^2 {\bf I}_2$, where $\sigma^2 = 4500$.
The fitted DINA CDM will be considered as the Data Generating Process CDM ($DGP_{CDM}$). To create a scenario in which the CDM is correctly specified, we refitted the initial original DINA CDM by sampling from the original fitted DINA CDM. Five different versions of the original DINA CDM, corresponding to "misspecified models," were used. These versions were created by flipping Tatsuoka's $\bf Q$-matrix elements with different probabilities (0\%, 1\%, 5\%, 10\%, 15\%, and 20\%). The 0\% misspecification level corresponds to the correctly specified case. For each level of $\bf Q$-matrix misspecification, we generate 5 different misspecified Q matrices at that level, allowing for variations in which elements are altered. We refer to a specific way of misspecifying the $\bf Q$-matrix at a given level as a "replication." Then, we fit each of the 5 replications of the $\bf Q$-matrix at a particular misspecification level to 50 bootstrap-generated datasets from the original DINA CDM. 

We computed  $GIMT_{Det}$ statistics as a function of misspecification level. Then we evaluated the performance of the $GIMT_{Det}$ in classifying correctly specified and misspecified models using the Receiver Operating Characteristic (ROC) curve. The ROC curve plots the true positive rate against the false positive rate at different decision thresholds, allowing for an investigation of the discrimination
performance of the $GIMT_{Det}$. Additionally, the area under the ROC curve (AUROC) can provide a quantitative measure of the $GIMT_{Det}$ discrimination, with a high AUROC value indicating that the $GIMT_{Det}$ effectively distinguishes between correctly specified and misspecified models. Conversely, an
AUROC close to 0.5 indicates no discrimination performance. 

\subsection{Results and Discussion}

\begin{figure} 
  \centering
  \includegraphics[width=50mm]{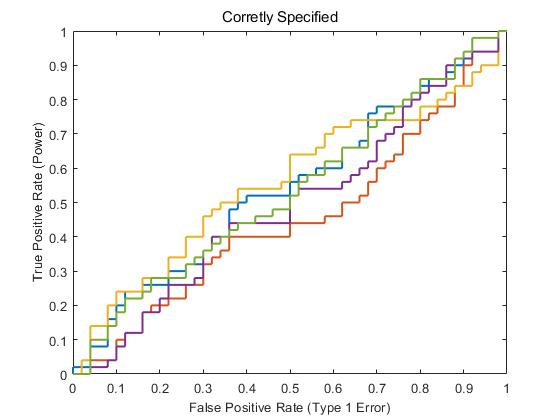}
  \includegraphics[width=50mm]{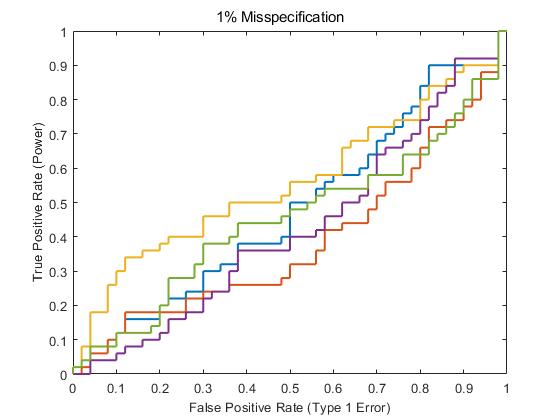}
  
  \includegraphics[width=50mm]{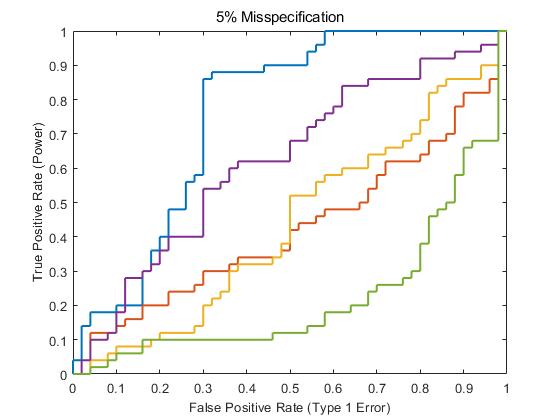}
  \includegraphics[width=50mm]{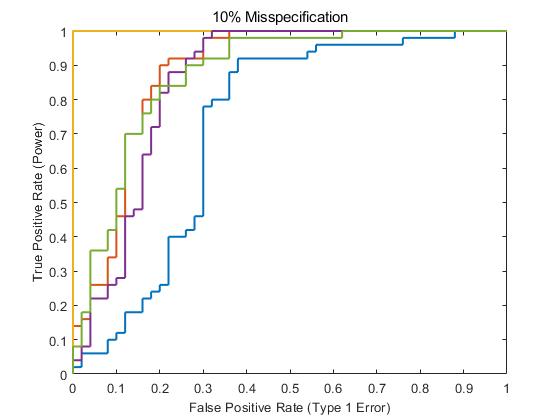}
  
  \includegraphics[width=50mm]{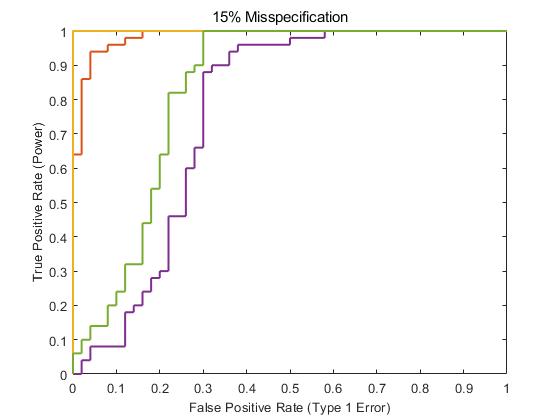}
  \includegraphics[width=50mm]{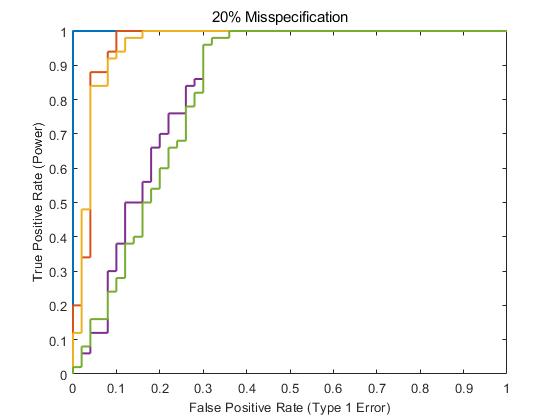}
  \caption{Influence of Misspecification Levels on ROC Curves: Level of misspecification: 0\%, 1\%, 5\%, 10\%, 15\%, and 20\%; Sample Size n = 536; Conducted with five replications for each level. 
The discrimination performance of $GIMT_{Det}$ was effective in the 10\%, 15\%, and 20\% cases, while there was no indication of discrimination performance in the 0\% and 1\% misspecification cases.}
  \label{fig:1}
\end{figure}

Table \ref{table:2} displays the AUROC values under various misspecification levels which shows that AUROC values increase
as the misspecification level increases. 
The results of the simulation study are also presented in Figure \ref{fig:1}, which displays how the ROC curves evolve at various misspecification levels (0\%, 1\%, 5\%, 10\%, 15\%, and 20\%) with a constant sample size of n=536.  At the 20\% misspecification level, the ROC curve demonstrates good discrimination performance, closely approaching the upper-left corner.

\par\null

\begin{table} 
  \centering
  \begin{tabular}{cccc}
    \toprule
    Misspecification level & AUROC Mean & \ Lower CI & \ Upper CI \\
    \midrule
    Correctly Specified & 0.51 & 0.46 & 0.57 \\
    1\% Misspecification & 0.46 & 0.38 & 0.54 \\
    5\% Misspecification & 0.54 & 0.23 & 0.85 \\
    10\% Misspecification & 0.87 & 0.75 & 0.99 \\
    15\% Misspecification & 0.91 & 0.78 & 1.00 \\
    20\% Misspecification & 0.92 & 0.82 & 1.00 \\
    \bottomrule
  \end{tabular}
  \caption{Average AUROC and Confidence Intervals at Different Levels of Model Misspecification, Conducted with five versions for Each Level.  The average AUROC increases with higher levels of misspecification. Notably, the 5\% misspecification case exhibits the most variation in the "version of misspecification," while reasonably good discrimination performance is consistently achieved for the 10\%, 15\%, and 20\% cases.}
  \label{table:2}
\end{table}

\par\null 
When the misspecification level was greater than 5\%, the GIMT showed good discrimination performance regardless of the manner in which the Q matrix was decimated. In addition, the GIMT did not detect model misspecification for the correctly specified case (0\%) and did not detect model misspecification for the slightly misspecified case (1\%). For moderate 
misspecification levels (5\%), however, only two of the five replications showed effective discrimination performance. There was also substantial variability
across replications for the 5\% case.
Currently, we view the 5\% level as a "transition region" and we plan to investigate this further using additional replications. In summary, the new GIMT method for misspecification detection in CDMs appears to be promising. Further research will continue to explore the capabilities and limitations of this approach.


\printbibliography 

\end{document}